\documentclass[a4paper]{article}

\usepackage{multicol}
\def\1{\c{c}}
\def\2{\c{C}}
\def\3{\.{I}}
\def\4{\"{a}}
\def\5{{\i}}
\def\6{$\beta$}
\def\7{\"{o}}
\def\8{\"{O}}
\def\9{\c{s}}
\def\0{\c{S}}
\def\*{\"{u}}
\def\;{\u{g}}
\def\:{\u{G}}

\usepackage{icrc2013}

\title{Studying the Supernova Remnant G31.9+0.0 in Gamma and X-Rays}

\shorttitle{Studying the Supernova Remnant G31.9+0.0 in Gamma and X-Rays}

\authors{
T\*l\*n Ergin$^{1}$,
Aytap Sezer$^{1}$,
Lab Saha$^{2}$,
Pratik Majumdar$^{2}$,
E. Nihal Ercan$^{3}$.
}

\afiliations{
$^1$ TUBITAK SpaceTechnologies Research Institute, Ankara, Turkey.\\
$^2$ Saha Institute of Nuclear Physics, Kolkata, India.\\
$^3$ Bogazici University, Physics Department, Istanbul, Turkey.\\
}

\email{tulun.ergin@tubitak.gov.tr}

\abstract{G31.9+0.0  (3C 391) is a Galactic mixed-morphology supernova remnant observed in GeV gamma rays by Fermi Gamma Ray Space Telescope's LAT (Fermi-LAT), as well as in the 0.3$-$10 keV X-ray band by Suzaku. In this paper, we will present the analysis results of X- and gamma-ray data of 3C 391 taken with Suzaku and Fermi-LAT. The X-ray spectrum of 3C 391 was fitted to a single-temperature variable abundance non-equilibrium ionization (VNEI) model with an electron temperature of kT$_e$ $\sim$ 0.57 keV, an absorbing column density of N$_{\rm{H}}$ $\sim$ 3.1$\times$10$^{22}$ cm$^{-2}$ and a very high ionization age  ($\tau$ $>$ 10$^{12}$ cm$^{-3}$ s), which suggest that the plasma has reached ionization equilibrium. The spectrum shows clearly detected emission lines of Mg, Si, and S. 3C 391 was detected in GeV gamma rays with a significance of $\sim$ 13 $\sigma$.  The spectrum was fitted with a log-parabola function, where the spectral index and beta parameters were found to be $\alpha$ = 2.35 $\pm$ 0.07 and $\beta$ = 0.366 $\pm$ 0.339. The integrated flux above 200 MeV was found as F =  (2.34 $\pm$ 0.37)$\times$10$^{-8}$ ph cm$^{-2}$ s$^{-1}$. These results are in agreement with the Fermi-LAT results given in the 2nd Fermi-LAT catalog. 
}

\keywords{Supernova Remnants, Molecular Clouds, Gamma Rays, X-Rays, Fermi-LAT, Suzaku.}

\begin{document}
\maketitle

\section{Introduction}
The Galactic supernova remnant (SNR) 3C 391 (G31.9+0.0), a member of the mixed-morphology class, has been observed in many different wavelengths, \cite{green2009}. The HI absorption measurements, \cite{radhakrishnan1972}, show that the distance to 3C 391 is at least 7.2 kpc (assuming a Galactocentric radius of 8.5 kpc) and for the emission without absorption indicate an upper limit of 11.4 kpc. 

In the radio band, 3C 391 is observed by VLA (\cite{reynoldsmoffet1993}) as a partial shell of 5$'$ radius with a breakout morphology, where the intensity of the radio emission in the shell rises in the bright northwest rim (NW) and drops and vanishes toward the southeast rim (SE). Breakout morphologies usually occur when the expanding SNR encounters a molecular cloud. Other indirect evidence for 3C 391 expanding into a medium with different gas density comes from X-rays. In the X-ray band, 3C 391 was observed with Einstein (\cite{wangseward1984}), ROSAT (\cite{rhopetre1996}), Chandra (\cite{chen2004}), and ASCA (\cite{chenslane2001}, \cite{kawasaki2005}). ROSAT and Einstein data reveal two bright X-ray peaks within the SNR: one with brightest X-ray peak interior to the weak SE rim and a fainter one in the interior of the bright NW radio shell. 

Using ROSAT observations, \cite{rhopetre1996} applied a single-temperature thermal model and obtained an absorbing column density of N$_{\rm{H}}$ $\sim$ 2.4$\times$10$^{22}$ cm$^{-2}$ and electron temperature of kT$_{e}$ = 0.5 keV. They also found enhanced abundances of Mg, Si, and S. \cite{chen2004} found that the X-ray spectra obtained from Chandra data can be best described by the VNEI model. The spectral fits showed that the diffuse emission has ionization parameter (n$_e$t) close to or higher than 10$^{12}$ cm$^{-3}$ s. This hints that the hot plasma in the remnant is very close to or in the ionization equilibrium. They found the electron temperature at $\sim$ 0.5$-$0.6 keV and estimated an age of  $\sim$ 4$\times$10$^3$ yr for the remnant. From the data of ASCA observation, \cite{kawasaki2005} found the electron temperature value as 0.53$^{+0.4} _{-0.3}$ keV by applying a NEI model to the spectra. They obtained an ionization timescale of $\tau \sim$2.5 ( $>$ 0.9 )$\times$10$^{12}$ cm$^{-3}$ s suggesting that the plasma has reached ionization equilibrium.

\cite{frail1996} observed two OH masers at 1720 MHz with velocities 105 and 110 km/s coincident with the southeast and northeast rim of 3C 391, respectively, showing first clear evidence for 3C 391 interacting with a molecular cloud. The CO(1-0) line observations of 3C 391 by \cite{wilner1998} showed that 3C 391 is embedded in the edge of a molecular cloud, supporting the evidence of SNR-molecular cloud interaction. Further evidence for shock interactions are the CS line observations by \cite{reachrho1999}, the measurements of strongly enhanced [O$_{\rm{I}}$] 63 $\mu$m (\cite{wilner1998}) at the NW rim of 3C 391, and the recent OH maser observations by \cite{hewitt2008}.

SNRs interacting with molecular clouds are interesting targets for the detection of gamma rays of hadronic origin. The hadronic mechanism producing gamma-ray emission can be explained as the decay of neutral pions, created in proton-proton interactions during the passage of SNR shocks through the dense molecular material, into two gamma rays. 3C 391 was observed in GeV gamma rays by Fermi Gamma Ray Space Telescope's LAT (Fermi-LAT) (\cite{atwood2009}) and it was listed in the 2nd Fermi-LAT  catalog (\cite{nolan2012}) as a point-source, called 2FGL J1849.3-0055. \cite{castroslane2010} analyzed the GeV data of 3C 391 and reported a 13 $\sigma$ detection. They showed that the peak of the significance map was shifted 4$'$ away from the northwestern edge of the radio shell. The spectrum of 3C 391 was best described as a power-law model with a spectral index of $\Gamma$ = $-$2.33 $\pm$ 0.11.  They found the integrated flux of 3C 391 as F(0.1$-$100 GeV) = (1.58 $\pm$ 0.26)$\times$10$^{-7}$ photons cm$^{-2}$ s$^{-1}$, \cite{castroslane2010}.  At TeV energies, H.E.S.S. reported integral flux upper limits at the 95\% CL in units of the flux of the Crab Nebula as F$^{UL}$ = 0.8 crab units, \cite{bochow2011}.

\section{Data Analysis and Results}
\subsection{X-Rays} 

3C 391 was observed with the X-ray Imaging Spectrometer (XIS) on board Suzaku on 2010 October 22, under the observation ID of 505007010 and an exposure time of 99.4 ks. Detailed descriptions of the Suzaku satellite, the XIS instrument, and the X-ray telescope are given in \cite{mitsuda2007}, \cite{koyama2007}, \cite{serlemitsos2007}, respectively. The XIS system consists of four CCD cameras (XIS 0, 1, 2, and 3). One of the cameras (XIS1) uses a back-illuminated (BI) CCD while the others (XIS0, 2, and 3) use front-illuminated (FI) CCDs. XIS2 has not been functional since an unexpected anomaly in 2006 November. The XIS was operated in the normal full-frame clocking mode. For the data reduction we used HEASoft package version 6.4.1. The latest calibration database (CALDB: 20130305) was used and fitting was carried out in the X-ray spectral fitting package (\emph{xspec}) version 11.3.2 (\cite{arnaud1996}). The redistribution matrix files (RMFs) of the XIS were produced by \emph{xisrmfgen}, and \emph{auxillary} response files (ARFs) were generated by \emph{xissimarfgen} (\cite{ishisaki2007}).
\begin{figure}[t]
\centering
\includegraphics[width=0.4\textwidth]{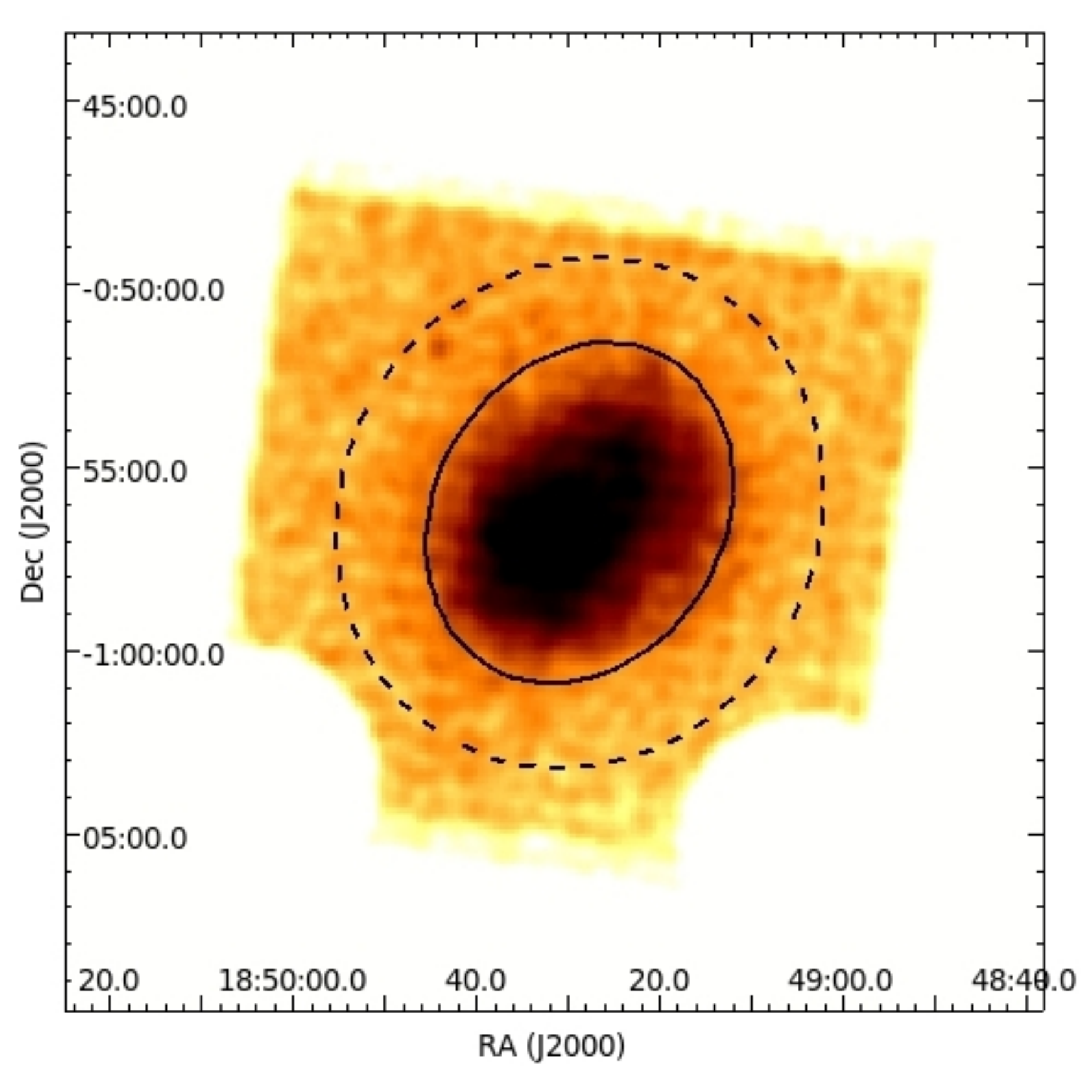}
\includegraphics[width=0.4\textwidth]{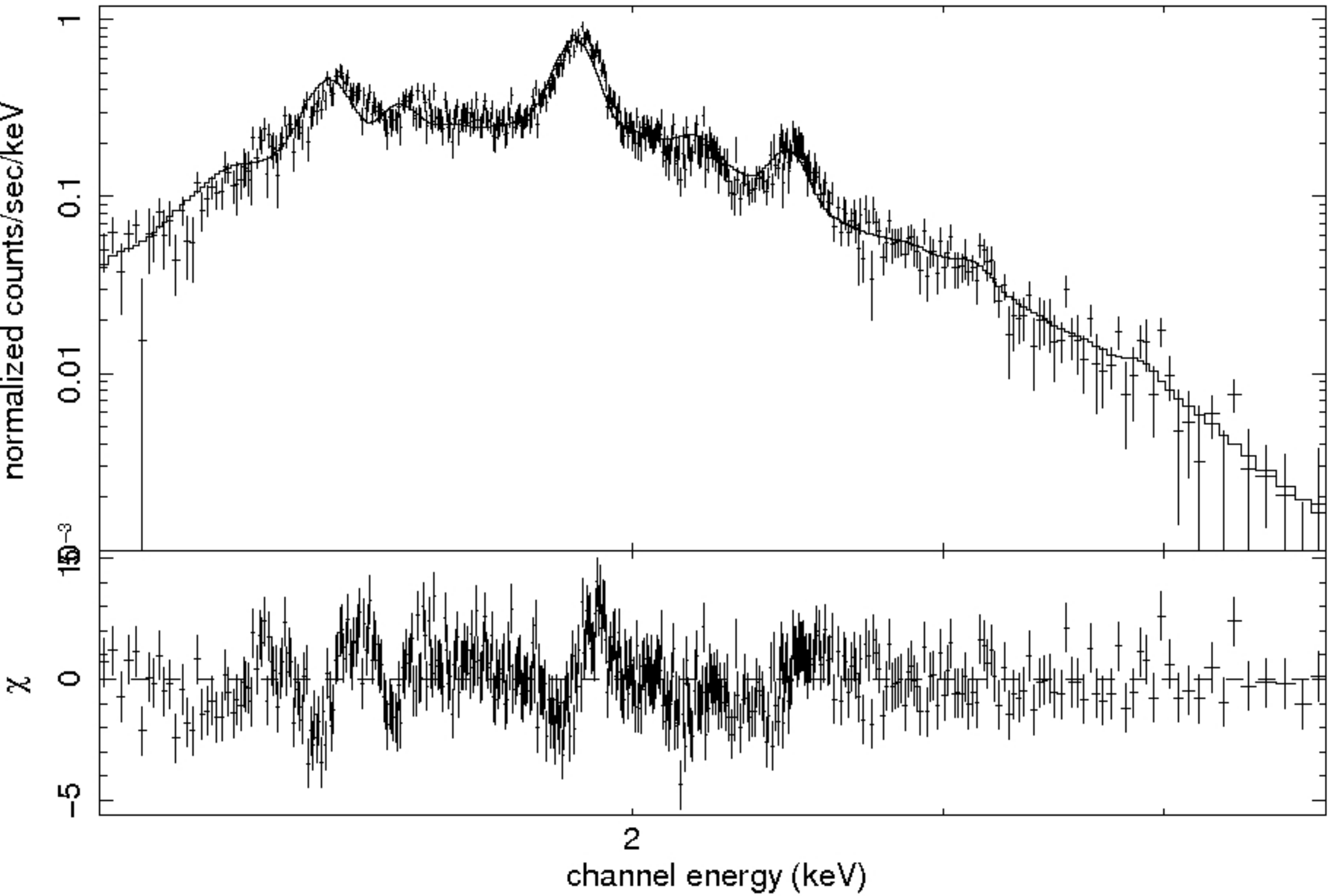}
\caption{\small{Upper Panel: Suzaku XIS1 image of 3C 391 in the 0.3$-$10.0 keV energy band. The source and background regions are shown as a solid ellipse and a dashed ellipse, respectively. The corners irradiated by the calibration sources of $^{55}$Fe are excluded. Bottom Panel: Background-subtracted XIS1 spectrum of 3C 391 in the 1.0$-$5.0 keV energy band fitted with an absorbed VNEI model. At the bottom of this panel, the residuals from the best-fitting model are shown.}}
\label{figure_1}
\end{figure}

At the top panel of Figure \ref{figure_1}, we present XIS1 image of the 3C 391 in the 0.3$-$10 keV energy band, where the source and background regions are shown as a solid and a dashed ellipses, respectively. The corners of the CCD chips illuminated by the $^{55}$Fe calibration sources were excluded.

The spectrum for the remnant was extracted from the elliptical source region (4.85$'$ $\times$ 3.94$'$) centered at RA(J2000) = 18h49m28.6s, Dec. (J2000) = $-$0$^{\circ}$ 56$'$ 16.4$''$ and background spectrum was extracted from the elliptical annulus surrounding the remnant (Figure \ref{figure_1}). The spectrum was binned to a minimum of 20 counts per bin using \emph{grppha} to allow use of the $\chi^2$ statistic. We tried various combinations of spectral models. An absorbed (wabs in xspec; \cite{morrison1983}) VNEI model, which is a model for a non-equilibrium ionization (NEI) collisional plasma with variable abundances
(\cite{borkowski2001}) gave the best $\chi^2$ value of 950.2/659 = 1.44. During model fitting N$_{\rm{H}}$, kT$_e$, n$_e$t, and the abundances of Mg, Si, and S were free parameters, while the other elemental abundances were fixed to their solar values (\cite{anders1989}). We fitted the FI and BI spectra simultaneously, but only the BI (XIS1) spectra are shown for simplicity. The background-subtracted XIS1 spectrum was fitted in the 1.0$-$5.0 keV energy band as shown in the bottom panel of Figure \ref{figure_1}. The best-fitting parameters (90\% confidence level) are presented in Table \ref{table_1}.
\begin{table}[h]
\begin{center}
\footnotesize{
\begin{tabular}{|l|c|}
\hline Parameter           					&		Value                                                  \\ \hline
	  N$_{\rm{H}}$ [10$^{22}$ cm$^{-2}$]	& 		3.1 $\pm$ 0.1			              \\ \hline
	  kT$_{e}$ [keV]					& 		0.57 $\pm$ 0.01			     \\ \hline		
	  Mg (solar) 						& 		1.3 $\pm$ 0.1			              \\ \hline		
	  Si (solar) 						& 		0.9 $\pm$ 0.1			              \\ \hline	
	  S (solar) 							& 		0.8 $\pm$ 0.1			              \\ \hline	
	  $\tau$ [10$^{12}$ cm$^{-3}$ s] 		& 	       11.7 $\pm$ 3.1			              \\ \hline	
	  Norm [ph cm$^{-2}$ s$^{-1}$] 		         & 		4.8 $\pm$ 0.3			              \\ \hline	
	  $\chi^2$/d.o.f 				 		& 		950.2/659 = 1.44 			     \\ \hline	
	   \end{tabular}
 }
\caption{ \small{Best-fitting spectral parameters of 3C 391 with corresponding errors at the 90\% confidence level in the 1.0$-$5.0 keV band.}}
\label{table_1}
\end{center}
\end{table}

\subsection{Gamma Rays}

The 3C 391 Fermi-LAT data was observed between 2008-08-04 and 2013-04-10. The events-data was taken from a region of interest (ROI) with a radius of 12$^{\circ}$ centered at the SNR's position of RA(J2000) = 282.343$^{\circ}$ and Dec(J2000) = $-$0.923$^{\circ}$ and the events suggested for Fermi-LAT Pass 7 for galactic point source analysis type were selected using \emph{gtselect} of Fermi Science Tools (FST) version v9r27p1. The Earth limb is a very bright source of gamma rays and during observations it comes very close to the edge of the field of view. To remove the contribution from this source, we cut out the gamma rays with reconstructed zenith angles greater than 105$^{\circ}$.
   
The gamma-ray events in the data were binned in energy at 15 logarithmic steps between 200 MeV and 300 GeV. For the binned likelihood analysis (\cite{abdo2009}), the matching energy dependent exposure maps were produced based on pointing direction, orientation, orbit location, and live-time accumulation of LAT. The large point-spread function (PSF) of LAT means that at low energies, source from outside the counts cube could affect the analyzed sources. To compensate for this and to ensure that the exposure map accounts for contributions from all the sources in the analysis region, exposure maps were created such that they included sources up to 10$^{\circ}$ outside the ROI. In addition, since at low energies the PSF is large, the exposure map should be expanded by another 10$^{\circ}$ to accommodate this additional exposure, \cite{abdo2009}. Since the exposure map uses square pixels, to match the binning in the counts cube, we generated an exposure map with 0.05$^{\circ}$$\times$0.05$^{\circ}$ bin size.
For the \emph{pointlike} analysis (\cite{kerr2011}), also based on the likelihood analysis (FST-v9r31p0), the radius of the analysis region was chosen as 2$^{\circ}$ to obtain comparable results to the ones obtained by the binned likelihood analysis. 

The spectral properties of the gamma-ray emission were studied by comparing the observation with models of possible sources in the ROI. Predictions were made by convolving the spatial distribution and spectrum of the source models with the instrument response function (IRF) and with the exposure of the observation. In the analysis we used the IRF version P7SOURCE$_{-}$V6. 

The model of the analysis region contains the diffuse background sources and all point-like sources from the 2nd Fermi-LAT catalog located at a distance $\leq$ 1.8$^{\circ}$ away from the center of the ROI. These point-like sources are shown in Table \ref{table_2}, where their positions, significances, and distances from the center of the ROI are given. The standard background model has two components: diffuse galactic emission (\emph{gal$_{-}$2yearp7v6$_{-}$v0.fits}) and isotropic component (\emph{iso$_{-}$p7v6source.txt}), which is a sum of the extragalactic background, unresolved sources, and instrumental background, where it's distribution is assumed to be isotropic. 

The background and source modeling was done by the binned likelihood analysis using \emph{gtlike} of FST. To determine the best set of spectral parameters of the fit, we vary the parameters until the maximum likelihood is maximized. In this analysis, we kept all the parameters of 2FGL J1847.2-0236 fixed, since it is relatively far away from the center of the ROI. For the rest of the point-like background sources, parameters except the normalizations were kept either fixed or free depending on the source strength and their distances from the ROI center. The detection of the source in this analysis is given by test statistics (TS) value, where larger TS values indicate that the null hypothesis (maximum likelihood value for a model without an additional source) is incorrect. This means that the source is present and its detection significance is approximately equal to the square root of TS.
\begin{table}[h]
\footnotesize{
\begin{tabular}{|l|p{0.7cm}|p{0.77cm}|p{0.9cm}|p{1.0cm}|}
\hline Source Name               & RA                & Dec            & Distance		& $\sqrt{TS}$ \\ 
                                                   & [$^{\circ}$]   & [$^{\circ}$] & [$^{\circ}$]	& [$\sigma$]\\ \hline
           2FGL J1849.3-0055    & 282.33	  & $-$0.917        &  0.02		& 8.99         \\ \hline
           2FGL J1849.9-0125c  & 282.49         & $-$1.426        &  0.53		& 7.46       \\ \hline
           2FGL J1850.7-0014c  & 282.69         & $-$0.248        & 0.76		& 9.18         \\ \hline
           2FGL J1848.2-0139c  & 282.07         & $-$1.651        & 0.78		& 11.36        \\ \hline
           2FGL J1847.2-0236    & 281.81         & $-$2.611        & 1.77		& 13.81         \\ \hline
    \end{tabular}
    }
\caption{ \small{The point-like sources from the 2nd Fermi-LAT source catalog included in the background model of 3C 391. The source at the top of the list represents 3C 391. The distances shown are from the center of the ROI. The significances are as given in the 2nd Fermi-LAT catalog. }}
\label{table_2}
\end{table}
\begin{figure}[t]
\centering
\includegraphics[width=0.5\textwidth]{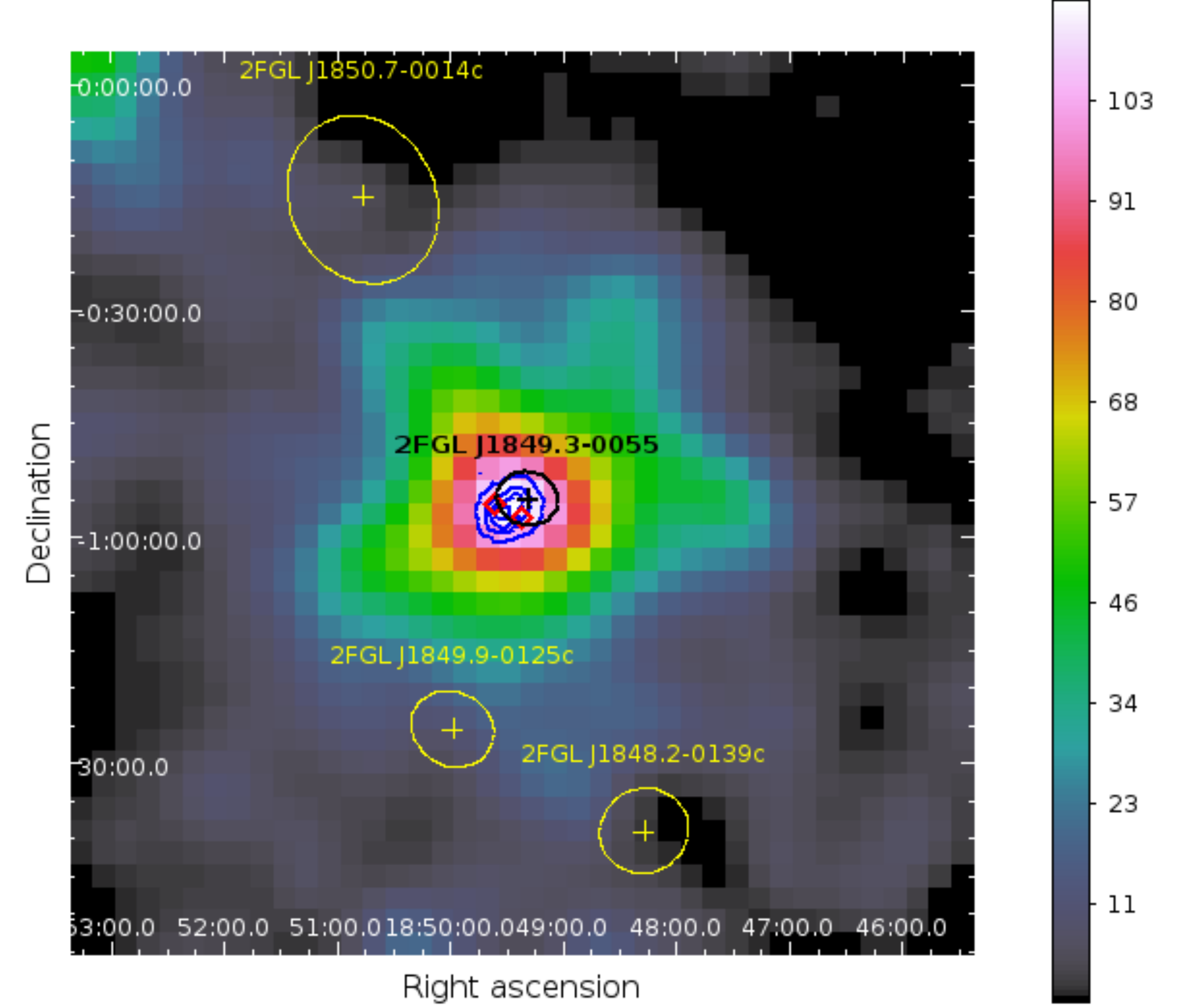}
\caption{\small{TS residual map produced by the Fermi-LAT analysis. The blue contours show the Suzaku data from Figure \ref{figure_1}, where 5 contours are used between 11 and 57 cts. The yellow crosses and circles represent the 2nd Fermi-LAT catalog sources and the black cross and circle is the GeV source from the 2nd Fermi-LAT catalog corresponding to SNR 3C 391. Two red diamonds represent the two masers detected by \cite{frail1996}.}}
\label{figure_2}
\end{figure}

Using \emph{pointlike} analysis we detected 3C 391 with a significance of $\sim$ 13 $\sigma$. We computed the best-fit position within the ROI of 3C 391, which was found as RA(J2000) = 282.360$^{\circ}$ and Dec(J2000) = $-$0.927$^{\circ}$, which enhanced the TS by 2.58 $\sigma$ over the position of 2FGL J1849.3-0055 in the 2nd Fermi-LAT catalog. Then the model was refitted using the best-fit position to compute the TS residual map and spectrum. Computing the TS map, we excluded 3C 391 from the model to see the full morphology of this SNR inside the ROI. Figure \ref{figure_2} shows the 2$^{\circ}$ $\times$ 2$^{\circ}$ TS residual map of 3C 391 and its nearby neighborhood with a bin size of 0.05$^{\circ}$$\times$0.05$^{\circ}$, where the blue contours represent the Suzaku background-subtracted X-ray data from the upper panel of Figure \ref{figure_1}, the yellow crosses and circles represent the 2nd Fermi-LAT catalog sources, and the black cross and circle is the GeV source from the 2nd Fermi-LAT catalog corresponding to SNR 3C 391. The peak value of the gamma-ray significance coincides with the X-ray remnant. The red diamonds indicate the OH maser locations reported by \cite{frail1996}.

To check the functional form of the 3C 391 spectrum, we first tried to fit a log-parabola and then a power-law function between 200 MeV and 300 GeV :  
\begin{itemize}
\item Log-parabola: \\
F(E)$^{LP}$ =  N$_{\circ}$ (E/E$_{b}$)$^{(\alpha + \beta \mbox{ln(E/E$_b$)})}$
\item Power-law:\\
 F(E)$^{PL}$ =  N$_{\circ}$ (E/E$_{\circ}$)$^{-\Gamma}$
\end{itemize}

The log-parabola fit resulted in the spectral index and beta parameter values of $\alpha$ = 2.35 $\pm$ 0.07 and $\beta$ = 0.366 $\pm$ 0.339, respectively, for a fixed E$_b$ value of 2430 MeV. These results were in good agreement with the results in the 2nd Fermi-LAT catalog (\cite{nolan2012}), where $\alpha$ = 2.35 $\pm$ 0.16 and $\beta$ = 0.308 $\pm$ 0.099. The power-law fit resulted in spectral index of $\Gamma$ = 2.28 $\pm$ 0.03, which is in agreement with the best-fit power-law index value given for 3C 391 in the 2nd Fermi-LAT catalog ($\sim$ 2.19), \cite{nolan2012}. This result also matches to the results obtained by \cite{castroslane2010},  $\Gamma$ = 2.33 $\pm$ 0.11. 

\begin{figure}[t]
\includegraphics[width=0.4\textwidth]{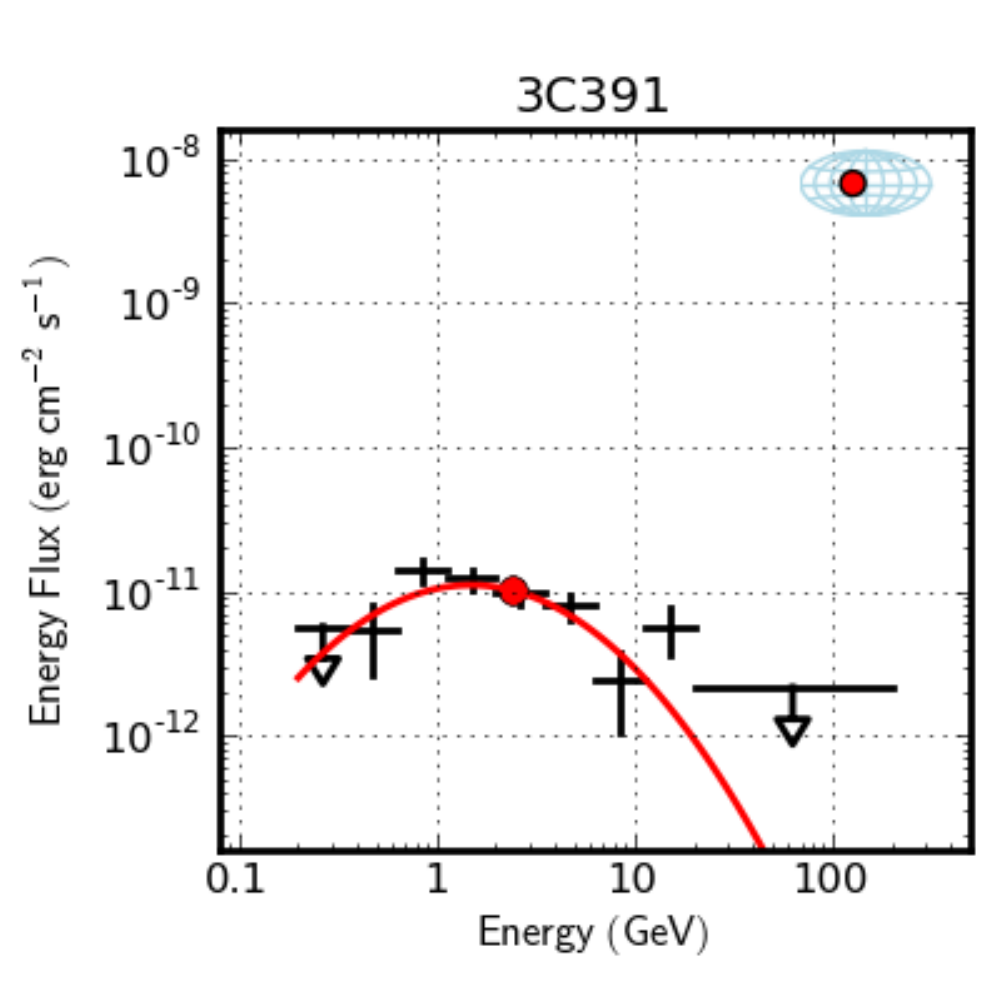}
\caption{\small{GeV gamma-ray spectrum of 3C 391 for the energy range of 0.2$-$300 GeV. }}
\label{figure_3}
\end{figure}

The integrated flux between 200 MeV and 300 GeV was found as F(E)$^{LP}$ =  (2.34 $\pm$ 0.37)$\times$10$^{-8}$ photons cm$^{-2}$ s$^{-1}$ and F(E)$^{PL}$ = (8.55 $\pm$ 0.22)$\times$10$^{-8}$ photons cm$^{-2}$ s$^{-1}$ for the log-parabola and power-law fits, respectively. The log-parabola fit gave a TS value of 160 for 3C 391, while the power-law fit resulted in a TS value of 146. The log-parabola fit gave a slightly higher TS value than the power-law fit, which was found to be consistent with the results in the 2nd Fermi-LAT source catalog. Figure \ref{figure_3} shows the GeV gamma-ray spectral data points and the statistical errors in black while the log-parabola fit is shown in red color.

\section{Conclusion}
In this work, we present results from a $\sim$ 99.4 ks observation of 3C 391 with Suzaku and from almost 4-years of observations of Fermi-LAT. The X-ray spectrum of the SNR is well described by the model of a thermal plasma with electron temperature of $\sim$ 0.57 keV, which is consistent with previous X-ray observations. We obtained a high absorbing column density of N$_{\rm{H}}~\sim$ 3.1 $\times$10$^{22}$ cm$^{-2}$, which is consistent with the previous X-ray observations implying that it is in the Galactic plane region. For full ionization equilibrium, the ionization timescale (n$_e$t), where n$_e$ and t are electron density and the time since the plasma was heated, should be $\ge$ 10$^{12}$ cm$^{-3}$ s (\cite{masai1984}). The VNEI model yields a very high ionization age ($\tau~>~$ 10$^{12}$ cm$^{-3}$ s), which suggests that the plasma has reached ionization equilibrium. This is consistent with the conclusion of the ASCA (\cite{chenslane2001}, \cite{kawasaki2005}) and Chandra (\cite{chen2004}) studies. We also have analyzed the GeV gamma-ray data of 3C 391 and detected the source with a significance of $\sim$ 13 $\sigma$, where the position of the peak TS value of the gamma-ray emission is coincident with the X-ray location. Since the PSF of Fermi-LAT is comparable or larger than the size of 3C 391, it is not possible to resolve the SNR morphology, therefore we could not determine if the gamma rays come form the NW rim of the SNR or from the direction of the maser spots on the opposite side of the shell. The spectra shows a log-parabola type shape with spectral index of $\alpha$ = 2.35 $\pm$ 0.07 and beta value of $\beta$ = 0.366 $\pm$ 0.339. The GeV flux above 200 MeV was found to be F(E)$^{LP}$ =  (2.34 $\pm$ 0.37)$\times$10$^{-8}$ photons cm$^{-2}$ s$^{-1}$. The same spectra was also fitted with a power-law model that gave the spectral index and flux as $\Gamma$ = 2.28 $\pm$ 0.032 and F(E)$^{PL}$ = (8.55 $\pm$ 0.22)$\times$10$^{-8}$ photons cm$^{-2}$ s$^{-1}$, respectively. Both models fit to the data successfully, but the log-parabola model is more favorable due to the higher significance value of the fit. The spectrum might be the result of hadronic interactions between 3C 391 and the associated molecular clouds in the vicinity. To understand if this is the case, we are going to do a detailed modeling of the spectrum to understand the possible emission mechanisms.


\begin{thebibliography}{}
\bibitem{abdo2009}  A.A. Abdo, et al., ApJS 183 (2009) 46 doi:10.1088/0067-0049/183/1/46. 
\bibitem{anders1989} E. Anders and N. Grevesse, Geochimica Cosmochimica Acta 53 (1989) 197 doi:10.1016/0016-7037(89)90286-X. 
\bibitem{arnaud1996} K.A. Arnaud, in G.H. Jacoby, J. Barnes, eds, ASP Conf. Ser. 101 (1996) 17. 
\bibitem{atwood2009} W.B. Atwood, et al., ApJ 697 (2009) 1071 doi:10.1088/0004-637X/697/2/1071.  
\bibitem{bochow2011} A. Bochow, Proc. of ICRC2011 7 (2011) 110 arXiv:1112.4976. 
\bibitem{borkowski2001} K.J. Borkowski, W.J. Lyerly, S.P. Reynolds,  ApJ 548 (2001) 820 doi:10.1086/319011. 
\bibitem{castroslane2010} D. Castro and P. Slane, ApJ 717 (2010) 372 doi:10.1088/0004-637X/717/1/372. 
\bibitem{chenslane2001} Y. Chen and P.O. Slane, ApJ 563 (2001) 202 doi:10.1086/323886. 
\bibitem{chen2004} Y. Chen, Y. Su, P.O. Slane, Q.D. Wang, ApJ 616 (2004) 885 doi:10.1086/425152. 
\bibitem{frail1996} D.A. Frail, W.M. Goss, E.M. Reynoso, E.B. Giacani, A.J. Green, R. Otrupcek, AJ 111 (1996) 1651 doi:10.1086/117904. 
\bibitem{gorham1996}  P.W. Gorham, P.S. Ray, S.B. Anderson, S.R. Kulkarni, T.A. Prince, ApJ 458 (1996) 257 doi:10.1086/176808. 
\bibitem{green2009}  D.A. Green, BASI 37 (2009) 45. 
\bibitem{hewitt2008} J.W. Hewitt, F. Yusef-Zadeh, M. Wardle, ApJ 683 (2008) 189 doi:10.1086/588652. 
\bibitem{ishisaki2007} Y. Ishisaki, et al., PASJ 59 (2007) 113.
\bibitem{kawasaki2005} M. Kawasaki, M. Ozaki, F. Nagase, H. Inoue, R. Petre, ApJ 631 (2005) 935 doi:10.1086/432591. 
\bibitem{kerr2011} M. Kerr, arxiv:1101.6072. 
\bibitem{koyama2007} K. Koyama, et al., PASJ 59 (2007) 23. 
\bibitem{masai1984} K. Masai, Ap\&SS 98 (1984) 367 doi:10.1007/BF00651415. 
\bibitem{mitsuda2007} K. Mitsuda, et al., PASJ 59 (2007) 1. 
\bibitem{moffetreynolds1994}  D.A. Moffett and S.P. Reynolds, ApJ 425 (1994) 668 doi:10.1086/174013.  
\bibitem{morrison1983} R. Morrison and  D. McCammon, ApJ 270 (1983) 119 doi:10.1086/161102. 
\bibitem{neufeld2007} D.A. Neufeld, et al., ApJ 664 (2007) 890 doi:10.1086/518857.
\bibitem{nolan2012} P.L. Nolan, et al., APJS 199 (2012) 31 doi:10.1088/0067-0049/199/2/31. 
\bibitem{radhakrishnan1972} V. Radhakrishnan, W.M. Goss, J.D. Murray, J.W. Brooks, ApJS 24 (1972) 49 doi:10.1086/190249.  
\bibitem{reachrho1999} W.T. Reach and J. Rho, ApJ 511 (1999) 836  doi:10.1086/306703.   
\bibitem{reach2002} W.T. Reach, J. Rho, T.H. Jarrett, P.-O. Lagage, ApJ 564 (2002) 302 doi:10.1086/324075.
\bibitem{reynoldsmoffet1993} S.P. Reynolds and D.A Moffett, AJ 105 (1993) 2226 doi:10.1086/116600.
\bibitem{rhopetre1996} J.-H. Rho and R. Petre, ApJ 467 (1996) 698 doi:10.1086/177645. 
\bibitem{serlemitsos2007} P.J. Serlemitsos, et al., PASJ 59 (2007) 9. 
\bibitem{wangseward1984} Z.R. Wang, F.D. Seward, ApJ 279 (1984) 705 doi:10.1086/161935.
\bibitem{wilner1998} D.J. Wilner, S.P. Reynolds, and D.A. Moffett, ApJ 115 (1998) 247 doi:10.1086/300190. 
\end{thebibliography}
\end{document}